\documentclass[secthm,seceqn,amsthm,ussrhead]{amsart}
\usepackage{amsmath,latexsym}
\usepackage[psamsfonts]{amssymb}
\usepackage{times}
\usepackage[mathcal]{euscript}
\numberwithin{equation}{section}

\newcommand{\bbT}{\mathbb T}
\newcommand{\bbZ}{\mathbb Z}

\renewcommand{\epsilon}{\varepsilon}

\newcommand{\be}{\begin{equation}}
\newcommand{\ee}{\end{equation}}

\newcommand{\R}{\mathbb{R}}

\newcommand{\T}{\mathbb{T}}

\newcommand{\Z}{\mathbb{Z}}

\newcommand{\cH}{{\mathcal H}}

{\bf}{\it}
\newtheorem{theorem}{Theorem}[section]
\newtheorem{lemma}[theorem]{Lemma}
\newtheorem{corollary}[theorem]{Corollary}

\newtheorem{definition}[theorem]{Definition}
\newtheorem{remark}[theorem]{Remark}
\newtheorem{proposition}[theorem]{Proposition}

\newtheorem{assumption}[theorem]{Assumption}

\date{\today}

\begin{document}

\title[Estimates on the number of eigenvalues of two-particle
discrete  Schr\"odinger ... ]{Estimates on the number of eigenvalues
of two-particle discrete  Schr\"odinger operators}

\author{Sergio  Albeverio $^{1,2,3}$}
\address{$^1$ Institut f\"{u}r Angewandte Mathematik,
Universit\"{a}t Bonn, Wegelerstr. 6, D-53115 Bonn\ (Germany)}

\address{
$^2$ \ SFB 256, \ Bonn, \ BiBoS, Bielefeld - Bonn}
\address{
$^3$ \ CERFIM, Locarno and Acc.Arch.USI (Switzerland) E-mail
albeverio@uni.bonn.de}
\author{Saidakhmat N. Lakaev$^{4,5}$}
\address{
{$^4$ Samarkand Division of Academy of sciences of Uzbekistan} \ {
E-mail:lakaev@yahoo.com; lakaev@wiener.iam.uni-bonn.de}}

\author{Janikul I.
  Abdullaev$^5$}
\address{$^5$ {Samarkand State University,University Boulevard 15, 7003004,Samarkand
(Uzbekistan)}  {E-mail: jabdullaev@mail.ru }}

\begin{abstract}
Two-particle discrete Schr\"{o}dinger operators $H(k)=H_{0}(k)-V$ on
the three-dimensional lattice $\Z^3,$ $k$ being the two-particle
quasi-momentum, are considered. An estimate for the number of the
eigenvalues lying outside of the band of $H_{0}(k)$ via the number
of eigenvalues of the potential operator $V$ bigger than the width
of the band of $H_{0}(k)$ is obtained. The existence of non negative
eigenvalues below the band of $H_{0}(k)$ is proven for nontrivial
values of the quasi-momentum $k\in \T^3\equiv (-\pi,\pi]^3$,
provided that the operator $H(0)$ has either a zero energy resonance
or a zero eigenvalue. It is shown that the operator $H(k), k\in
\T^3,$ has infinitely many eigenvalues accumulating at the bottom of
the band from below if one of the coordinates $k^{(j)},j=1,2,3,$ of
$k\in \T^3$ is $\pi.$
\end{abstract}

\maketitle



\section{Introduction}

The study of the low-lying excitation spectrum for  lattice
Hamiltonians of systems with an infinite number of degrees of
freedom has recently attracted considerable attention (see, e.g.,
\cite{KoMi, MiSu, Zhi}).

See also \cite{ALzMahp04,ALMzMma05,FIC,GrSh,Mat,Mog,RSIV,Yaf00} for
general expositions and the discussion of particular problems of the
theory of discrete Schr\"{o}dinger operators on lattices, including
applications to solid state physics.

The main aim of the present paper is to provide a thorough analysis
of the dependence of the number of eigenvalues of a two-particle
lattice Schr\"odinger operator with  emphasis on its non-trivial
dependence on the total quasi-momentum and threshold phenomena.

In the lattice case, similarly as in the continuous case, the
two-body problem reduces to a one-body problem through the usual
split-off of the center of mass and the  introduction of the
relative coordinates. As a result, the two-body energy operator is
represented as a direct von Neumann integral, where the fiber
Hamiltonian $H(k) $ depends parametrically on the internal binding
the quasi-momentum $k\in \T^3\equiv (-\pi,\pi]^3$ (see
\cite{ALzMahp04,ALMzMma05,FIC,GrSh,Lfa93,RSIV,Yaf00}).

One can observe that the number of eigenvalues of the self-adjoint
perturbation $A-V$ of an abstract bounded self-adjoint operator $A$
in the Hilbert space $\cH$  depends on the {\it{width}} of the
spectrum of the non-perturbed operator $A$ (see Definition
\ref{defin.3.1} below for this concept).

In the present paper we obtain  an estimate for  the number of
eigenvalues of the perturbed operator $A-V$ lying outside  of  the
essential spectrum $\sigma_ {\text{ess}}(A-V)$
 depending on the number $n_{+}(w_{s}(A),|V|)$ of
eigenvalues of the perturbation operator $V,$ which are larger in
absolute value than the width of the essential spectrum  $w_{s}(A)$
of the operator $A$ (Theorem \ref{number}).

As a consequence of these results we obtain an estimate for the
number of eigenvalues of  the two-particle discrete
Schr\"{o}dinger operator $H(k)=H_{0}(k)-V$ via the number of
eigenvalues of the interaction operator $V$ which are bigger than
the width $w_b (H_{0}(k))$ of the band spectrum (band) of $
H_{0}(k)$ (Theorem \ref{neraven}) .

In the case of continuous Schr\"odinger operators one observes the
emergence of negative eigenvalues from the bottom of the continuous
spectrum.

 This
phenomenon is closely related to the existence of zero-energy
resonances and zero eigenvalues at the bottom of continuous spectrum
of the two-particle Schr\"odinger operators (so-called critical
Hamiltonians).

The appearance of negative eigenvalues for critical (non-negative)
Schr\"odinger operators under infinitesimally small negative
perturbations is remarkable: the presence of a zero-energy
resonances in at least two of the two-particle operators
(subsystems) leads to the existence of infinitely many eigenvalues
accumulating to the bottom of the three-particle continuum for the
corresponding three-particle Schr\"odinger operator (the Efimov
effect)
  (see, e.g., \cite{AHW,ALzMahp04,Lfa93,OvSi,Sob,
 Tam91, Tam93, Yaf74}).

The presence of eigenvalues below the bottom of the continuous
spectrum is especially remarkable for the three-particle lattice
Schr\"odinger operators (Hamiltonians) $H_3(K),K\in\T^3$: the
presence in at least one of the two-particle operators
(subhamiltonians) implies for $H_3(K)$ the finiteness of the number
of eigenvalues for all sufficiently small nonzero values of the
three-particle quasi-momentum $K\in\T^3$ even in the case, where all
two-particle subhamiltonians have a zero energy resonance (see
\cite{ALzMahp04}).

If we compare the results in the lattice case with those in the
continuous case  we see that in the former case   there is a
mechanism, which causes the emergence of eigenvalues  from the
threshold of the Hamiltonians, this mechanism has nothing to do with
additional (effectively negative) perturbations of the potential
term. In fact the mechanism is provided by the nontrivial dependence
of the Hamiltonians $H(k)$ on the quasi-momentum $k\in\T^3.$ We
prove this fact using the lattice analogue of Birman-Schwinger
principle (Theorem \ref{existence}).

The number of eigenvalues outside of the band of the two-particle
Schr\"{o}dinger operators essentially depends on the
 {\it{ width  $w_{jb}(\cdot)$ of the band in the
direction $j=1,2,3$}}, ($e^j$ being the unit vector along the j-th
direction in $\Z^3$).

We establish that the operator $H(k)$ has infinitely many positive
eigenvalues lying outside of the band spectrum if {\it{ width
$w_{jb}(\cdot)$ of the band in  direction $j$}} vanishes for some
$j=1,2,3$ (Theorem \ref{cheksiz}). This corresponds to having an
effective mass in direction $j,j=1,2,3,$ which increases to $+
\infty$

The paper is organized as follows.

In Section 2 we describe  the two-particle Hamiltonians  in the
coordinate representation and two-particle discrete Schr\"odinger
operators $H(k)$ in the momentum representation. In Section 3 we
obtain  an estimate for  the number of eigenvalues of the perturbed
operator $A-V$ lying outside its essential spectrum by the number of
$V$. In Section 4 we introduce the notions {"\it{width of the
band}"} and {"\it{width of the band in the direction j}"} for the
two-particle discrete Schr\"{o}dinger operators $H(k)=H_{0}(k)-V$
and prove Theorem \ref{neraven}. In Section 5 we prove the
Birman-Schwinger principle for the two-particle discrete
Schr\"{o}dinger operators $H(k).$ In Section 6 we introduce the
notion of a zero-energy resonance for two-particle discrete
Schr\"{o}dinger operators $H(k)$ and prove Theorems \ref{existence}
and \ref{cheksiz}.

\section{ Energy Operators of two particles on a lattice in the coordinate
representation and the two-particle discrete Schr\"{o}dinger
operators}

The free Hamiltonian $\hat H_{0}$ of the system of two quantum
particles  on the three-dimensional lattice $\bbZ^3$ is defined by
the following (bounded) self-adjoint operator on the Hilbert space
$\ell^2((\bbZ^3)^2)$

\begin{equation*}\label{free0}
\hat H_{0}=\frac{1}{2m_1}\Delta_{x_1}+\frac{1}{2m_2}\Delta_{x_2},
\end{equation*}
with $\Delta_{x_1}={\Delta}\otimes I$ and
$\Delta_{x_2}=I\otimes\Delta$, where $m_\alpha>0$  denotes the
mass of the particle $\alpha$ and $I$ is the identity operator on
$\ell^2({\Z}^3).$

The Laplacian $\Delta$ is a difference operator which describes
the transport of a particle from a site to the nearest neighboring
site, i.e.,
 $$
 (\Delta\hat{\psi})(x)= \sum_{\mid s\mid =1}
[\hat{\psi}(x)- \hat{\psi}(x+s)],\quad
\hat{\psi}\in\ell^2({\Z}^3).
$$

The total two-particle Hamiltonian $\hat H$ on the Hilbert space
$\ell^2(({\Z}^3)^2)$ describes the interaction between
 two particles
\begin{equation*}
\hat H =\hat H_{0}- \hat{V},
\end{equation*}
where
\begin{equation*}
(\hat{V}\hat{\psi})(x_1,x_2) ={\hat{v}(x_1 -x_2
)\hat{\psi}(x_1,x_2)},\quad \hat{\psi} \in \ell^2(({\Z}^3)^2),
\end{equation*}
and $\hat v$ is a real bounded function on $\Z^3.$

\begin{assumption}\label{hypoth}
Assume that $\hat{v}(s)$ is even function on $\Z^3$  verifying
\begin{equation*}\lim _{|s|\to
 \infty}|s|^{3+\kappa}\hat{v}(s)=0,\quad \kappa>1/2.
 \end{equation*}
\end{assumption}

Under above assumptions the two-particle Hamiltonian $\hat{H}$ is
a well defined bounded self-adjoint operator on $
\ell^2(({\Z}^3)^2).$

Recall that the study of spectral properties of the two-particle
Hamiltonian $\hat{H}$ in the momentum representation reduces to
the spectral analysis of the {\it{two-particle discrete
Schr\"{o}dinger operators }} $H(k),\,\, k \in {\bbT}^3,$ acting in
the Hilbert space $L^2(\T^3)$ (see
\cite{ALzMahp04,Lfa93,ALMzMma05}
\begin{equation}\label{two-part}
H(k) =H_{0} (k)-V.
\end{equation}

 The non-perturbed operator
$H_{0}(k)$ is the multiplication operator by the function $E_k(
q)$ on $L^2(\T^3)$
\begin{equation*}
(H_{0}(k)f)(q)=E_k( q)f(q),\quad f \in L^2(\T^3),
\end{equation*} where
 \begin{equation*}
E_{k}(q)=\frac{1}{m_1} \varepsilon (\frac{1}{2}k+q)
+\frac{1}{m_2}\varepsilon (\frac{1}{2}k-q),
\end{equation*}
 $$\varepsilon(q)=\sum_{j=1}^3(1- cos q^{(j)}),\quad
q=(q^{(1)},q^{(2)},q^{(3)}) \in \T^3.
$$

 The interaction (perturbation) operator $V$ acts on  the Hilbert space
 $L^2({\T}^3)$ by
$$ (V f)(q)=(2\pi )^{-3/2}\int\limits_{\T^3} v(q-s)f(s)d s, \quad
f \in L^2(\T^3). $$

Here the kernel function $v$ is given by the Fourier series
\begin{equation*}
v (q)=(2\pi )^{-3/2}\sum_{s\in {{{\Z^3}}}}  \hat v
(s)\,e^{\mathrm{i}(q,s)},
\end{equation*}
with $$ (q,s)={\sum }_{j=1}^3 q^{(j)}s^{(j)}, \quad
q=(q^{(1)},q^{(2)},q^{(3)})\in {\T}^3,\quad
s=(s^{(1)},s^{(2)},s^{(3)})\in {{\Z}}^3. $$

 We see that the operator $H(k), k\in \T^3,$ does not split into the
sum of a center-of-mass term and a relative kinetic energy term
contrary to the continuum case, where one as is center-of-mass.

\section{On the  number of eigenvalues of  bounded self-adjoint
operators}

We establish some auxiliary results on the  number of eigenvalues
for an abstract operator on a Hilbert space ${\cH}$.

 For  a bounded self-adjoint operator $A$ in a Hilbert space
${\cH},$ we define $n_{+}(\mu,A)$ (resp. $n_{-}(\mu,A)$) as
$$
n_{+}(\mu,A)=sup\{ \dim{L}: L\subset{\cH};(Af,f)>\mu,f\in L,
||f||=1 \}$$
 resp. $$n_{-}(\mu,A)=sup\{ \dim{L}: L\subset{\cH};(Af,f)<\mu,f\in L,
 ||f||=1\}$$
 The value  $n_{+}(\mu,A)$ (resp. $n_{-}(\mu,A)$) is equal to the
 infinity if $\mu$ is in the essential spectrum and $n_{+}(\mu,A)$ (resp.
 $n_{-}(\mu,A)$)
 is finite, it is equal to the number of the eigenvalues of $A$ bigger (resp.
 smaller)
 than $\mu.$

Now we introduce the  notion of "{\it{ width  of the spectrum of
the self-adjoint operator $A$}}".

\begin{definition}\label{defin.3.1}
The width of the operator $A,$ denoted by $w_{s}(A)$, is defined
to be
$$w_{s}(A)=M(A)-m(A),$$
 where
 $M(A)=\sup_{\|f\|=1} (Af,f)$ and
$m(A)=\inf_{\|f\|=1} (Af,f).
$
\end{definition}
\begin{theorem}\label{number} Let $A$ be a bounded self-adjoint
operator and $V$ be a compact self-adjoint operator in a Hilbert
space ${\cH}.$ Then
\begin{equation*}
n_{-}(m(A),A-V)\geq n_{+}(w_{s}(A),V)\quad\big(\mbox{resp.} \quad
n_{+}(M(A),A-V)\geq n_{-}(-w_{s}(A),V)\big) .
\end{equation*}
\end{theorem}
\begin{proof} Let $A$ be a scalar operator, i.e., $A=\mu I,\mu\in\R^1.$
Then  $w_{s}(A)=0$ and hence we have
 $$ n_{-}(\mu,A-V)=
n_{+}(0,V). $$

Now let $A$ be a  non-scalar operator, i.e.,  $A\neq \mu I.$ Then
$ w_{s}(A)>0.$ Since $V$ is a compact operator the number
$n_{+}(w_{s}(A),V)$ is finite. Let $\cH_{(w_{s}(A),+\infty)}(V) $
be the subspace generated by the eigenfunctions of $V$ associated
with the eigenvalues bigger than $w_{s}(A).$ Then
$n_{+}(w_{s}(A),V)=\dim \cH_{(w_{s}(A),+\infty)}(V)$ and for any
$f \in \cH_{(w_{s}(A),+\infty)}(V), ||f||=1$
 the following relations  $$
(Af,f)-(Vf,f)< M(A)-w_{s}(A)=m(A) $$ hold.
 This implies
that $$n_{-}(m(A),A-V)\geq \dim \cH_{(w_{s}(A),+\infty)}(V) =
n_{+}(w_{s}(A),V).$$ Similarly we get $$n_{+}(M(A),A-V)\geq
n_{-}(-w_{s}(A),V).$$
\end{proof}
\begin{corollary}\label{sum}
Let $A$ be a bounded self-adjoint operator and $V$ be compact
self-adjoint operator in  ${\cH}.$ Then
\begin{equation*}
n_{-}(m(A),A-V)+n_{+}(M(A),A-V)\geq n_{+}(w_{s}(A),|V|).
\end{equation*}
\end{corollary}
\begin{proof}
It is sufficient to notice that the singular values of a compact
self-adjoint operator $V$ are precisely the eigenvalues of $|V|.$
\end{proof}

\section{On the number of eigenvalues of the
two-particle discrete Schr\"odinger operators $H(k)$,
$k\in {\bbT}^3$ }

Under Assumption \ref{hypoth} the perturbation $V$ of the operator
$H_0(k)$ is a trace  class operator and therefore in accordance
with the well known invariance of the absolutely continuous
spectrum under trace class perturbations  the absolutely
continuous spectrum of the operator $H(k)$ defined by
\eqref{two-part} fills in the following interval on the real axis:
$$ \sigma_{\text{ac}}(H(k))= [{ E}_{\min}(k) ,{E }_{\max}(k)], $$
where
 $$
{E }_{\min}(k)=\min_{q{\in }{\bbT}^3}E_{k} (q),\quad {E }_{\max}
(k)=\max_{q{\in }{\bbT}^3} E_{k} (q).
 $$
Now we will apply the results obtained in Section 3 to the
two-particle discrete Schr\"odinger  $H(k)$. We recall that the
operator $H_0(k)$ has a band.

Let us introduce the notion "{\it{ width of the band of
$H_{0}(k).$}}"
\begin{definition} The {\it{width of the band of $H_0(k)$}}, denoted
by $w_b(H_0(k))$, is defined to be $$ w_b(H_0(k))\equiv \max_{q\in
\T^3}E_k(q)-\min_{q\in \T^3}E_k(q). $$
\end{definition}

\begin{theorem}\label{neraven} (i) For any $k\in \T^3$ the
following inequality holds $$ n_{-}(E_{\min} (k),H(k))\geq n_{+}(
w_b(H_0(k)),V).$$ (ii) Let $m_{1}=m_{2}$ and
$k=\tilde{\pi}=(\pi,\pi,\pi)$. Then
 the equalities  hold
$$
n_{-}(\frac{6}{m},H(\tilde{\pi}))=n_{+}(0,V)\quad\mbox{and}\quad
n_{+}(\frac{6}{m},H(\tilde{\pi}))=n_{-}(0,V) .$$
\end{theorem}
\begin{proof}
The  part $(i)$ of Theorem \ref{neraven} is a consequence of
Theorem \ref{number}. From the condition $(ii)$ of Theorem
\ref{neraven} we conclude that $E_{k}(p)=6(m)^{-1}$,  hence that
$w_b(H_0(\tilde \pi))=0,$ and finally  that  $H_0(\tilde
\pi)=6(m)^{-1}I$ is a scalar operator. Thus  the numbers
  $\{\frac{6}{m}- \hat v(x), x\in
\Z^3\}$ are eigenvalues of the operator $H(\tilde{ \pi} ).$
 \end{proof}
The Corollary \ref{sum} yields immediately the following
\begin{corollary}
For any $k\in \T^3$ the following inequality
\begin{equation*}
n_{-}(E_{\min}(k),H(k))+n_{+}(E_{\max}(k),H(k))\geq n_{+}(
w_b(H_0(k)),|V|)
\end{equation*}
holds.
\end{corollary}
\qed

 Now we introduce the notion "{\it{ width of the  band spectrum of
$H_0(k)$ in the direction $ j=1,2,3$}}".

\begin{definition}\label{def.4.4} Let $k=(k^{(1)},k^{(2)},k^{(3)})\in \T^3.$
By the  width $w_{jb}(k^{(j)})$ of the band spectrum in the
direction $ j=1,2,3$ of $H_0(k)$ we mean
\begin{equation*}
w_{jb}(k^{(j)})\equiv \max_{q^{(j)}\in (-\pi,\pi]} E_{k}(q)-
\min_{q^{(j)}\in (-\pi,\pi]} E_{k}(q).
\end{equation*}
\end{definition}
\begin{lemma}\label{width}
The width of the band spectrum in the direction
$w_{jb}(k^{(j)}),\,\, j=1,2,3,$ depends only on $k^{(j)}\in
(-\pi,\pi]$ and the equality
\begin{equation*}
w_b(H_0(k))=\sum_{j=1}^3w_{jb}(k^{(j)})
\end{equation*} holds.
\end{lemma}
\begin{proof} The function $E_k(p)$ can be rewritten as
\begin{equation}\label{sodda}
 E_k(p)=3(\frac{1}{m_1}+\frac{1}{m_2})-
 \sum_{j=1}^3(a(k^{(j)})\cos p^{(j)}+b(k^{(j)})\sin p^{(j)}),
\end{equation}
where the coefficients $a(k^{(j)})$ and $b(k^{(j)})$ are given by
\begin{equation*}
a(k^{(j)})=(\frac{1}{m_1}+ \frac{1}{m_2})
\cos\frac{k^{(j)}}{2},\quad b(k^{(j)})=(\frac{1}{m_1}-
\frac{1}{m_2})\sin\frac{k^{(j)}}{2}.
\end{equation*}
 The equality \eqref{sodda} implies the following
representation for $E_k(p)$
\begin{equation}\label{represent}
E_k(p)=3(\frac{1}{m_1}+\frac{1}{m_2})-\sum_{j=1}^3r(k^{(j)})\cos
(p^{(j)}- p(k^{(j)})),
\end{equation}
where
\begin{equation*}
r(k^{(j)})=\sqrt{a^2(k^{(j)})+b^2(k^{(j)})},\quad
p(k^{(j)})=\arcsin\frac{b(k^{(j)})}{r(k^{(j)})},\quad k^{(j)}\in
(-\pi,\pi].\end{equation*}
 From Definition \ref{def.4.4} of the {\it width $w_{jb}(k^{(j)})$
 of the band in the direction
 $j=1,2,3$}
 and the representation \eqref{represent} for $E_k(p)$ we
conclude that $$ w_{jb}(k^{(j)})=2 r(k^{(j)})\quad \text{and}\quad
w_{b}(H_0(k))=\sum_{j=1}^3 2 r(k^{(j)}).$$
 \end{proof}

The proof of Lemma  4.5 implies  the following
\begin{corollary}
(i) The functions $w_{jb}(k^{(j)})\equiv 2r(k^{(j)}),j=1,2,3,$ are
real analytic, even and positive on the interval $(-\pi,\pi).$

(ii) The equality $$w_{jb}(k^{(j)})=0$$ holds  if and only if
$m_1=m_2=m$ and $k^{(j)}=\pi$ for some $j=1,2,3.$
\end{corollary}

\section {The Birman-Schwinger principle for the
two-particle discrete Schr\"odinger operators $H(k)$, $k\in \T^3$}

\begin{assumption}
Assume that the interaction operator $V$ is positive.
\end{assumption}

 Since the operator $\hat V$ is  the multiplication
operator by the  positive function $\hat v(s)$ its positive root
$\hat V^{\frac{1}{2}}$ is the multiplication operator by $\hat
v^{\frac{1}{2}}(s)$. Hence  the positive root $V^{\frac{1}{2}}$ of
the positive operator $V$ has form
\begin{equation*}
(V^{\frac{1}{2}}f)(p)= \frac{1}{(2\pi)^{\frac{3}{2}}}
\int_{\T^3}v^{\frac{1}{2}}(p-p')f(p')dp',
\end{equation*}
where the kernel function $v^{\frac{1}{2}}(p)$ is  the inverse
Fourier transform of the function $\hat
v^{\frac{1}{2}}(s)$,\,i.e.,
\begin{equation*}
v^{\frac{1}{2}}(p)=\frac{1}{(2\pi)^{\frac{3}{2}}}\sum_{s \in
\Z^3}\hat v^{\frac{1}{2}}(s) e^{\mathrm{i}(p,s)}.
\end{equation*}

We define for any $k\in (-\pi,\pi)^3$ and $z\leq
E_{\text{min}}(k)$  \big (and also for any $k\in \T^3\setminus
(-\pi,\pi)^3$ and $z< E_{\text{min}}(k)$ \big)
 the integral operator $G(k,z)$ resp.
$G_\frac{1}{2}(k,z)$ with the kernel $G(k,z;p,q)$ resp.
$G_\frac{1}{2}(k,z;p,q)$
\begin{equation}\label{yadroG}
G(k,z;p,q)=
\frac{1}{(2\pi)^{3}}\int\limits_{{\T}^3}v^\frac{1}{2}(p-t)
(E_k(t)-z)^{-1}v^\frac{1}{2}(t-q) d t
\end{equation}
resp.
\begin{equation*}
G_\frac{1}{2}(k,z;p,q)=\frac{1}{(2\pi)^{\frac{3}{2}}}
v^\frac{1}{2}(p-q)(E_k(q)-z)^{-\frac{1}{2}}.
\end{equation*}

The proof of the following variant of the Birman-Schwinger
principle for two-particle discrete Schr\"odunger operator  $H(k)$
follows similar lines as in the case of quantum   particles moving
on $\R^3.$

\begin{lemma}\label{l.b-s} For any $k\in \T^3$ and $z<E_{\text{min}}(k)$
 the
operator $G(k,z)$ acts on $ L^2({\T}^3),$ is positive and the
equality
\begin{equation}\label{b-s=}
n_{-}(z,H(k)) =n_{+}(1,G(k,z))\end{equation} holds.
\end{lemma}

Now we are going to obtain a generalization of the
Birman-Schwinger principle for the two-particle Schr\"odinger
operators on the lattice $\Z^3$.

\begin{theorem}\label{gl.b-s}
 For any $k\in (-\pi,\pi)^3$  the
operator $G(k, E_{\text{min}}(k))$ acts on $ L^2({\T}^3),$ is
positive, trace class and the equality
\begin{equation*}
n_{-}( E_{\text{min}}(k),H(k)) =n_{+}(1,G(k,
E_{\text{min}}(k)))
\end{equation*} holds.
\end{theorem}
\begin{proof} For any $z \leq E_{\text{min}}(k)$ the operator
$G_\frac{1}{2}(k,z)$ is Hilbert-Schmidt. Since the equality
$$G(k,z)=G_\frac{1}{2}(k,z)(G_\frac{1}{2}(k,z))^*$$ holds the
operator $G(k,z)$ is positive and belongs to the trace class. It
is easy to show that there exists $C>0$ so that the inequality
$$||G(k,E_{\text{min}}(k))-G(k,z)||\leq C
\sqrt{E_{\text{min}}(k)-z} $$ holds.

Let us show that
$n_{-}(E_{\text{min}}(k),H(k))=n_{+}(1,G(k,E_{\text{min}}(k))).$
 Since $G(k,E_{\text{min}}(k))$ is
a compact operator the number $n_{+}(1,G(k,E_{\text{min}}(k)))$ is
finite. For any $\psi \in L^2(\T^3)$ and $z< E_{\text{min}}(k)$
the following relations $$(G(k,z)\psi,\psi)=
\int\limits_{{\T}^3}\frac{|V^{1/2}\psi(p)|^2 d p}{E_k(p)-z}\leq
\int\limits_{{\T}^3}\frac{|V^{1/2}\psi(p)|^2 d
p}{E_k(p)-E_{\text{min}}(k)}=(G(k,E_{\text{min}}(k))\psi,\psi) $$
hold.

Consequently we have
\begin{equation}\label{monoton}
n_{+}(1,G(k,z))\leq n_{+}(1,G(k,E_{\text{min}}(k))).
\end{equation}
Let $\lambda_{1}\geq\lambda_{2}\geq\ldots\geq\lambda_{n}>1$ be the
eigenvalues  and $\psi_{1},\psi_{2},\ldots,\psi_{n}$ the
corresponding eigenfunctions of the operator
$G(k,E_{\text{min}}(k)).$ Denote by ${\cH}_n$ the subspace
generated by
 the eigenfunctions $\psi_{1},\psi_{2},\ldots,\psi_{n}.$ Since for all
$z\in
U_{\delta}(E_{\text{min}}(k))=(E_{\text{min}}(k)-\delta,E_{\text{min}}(k)],\delta>0$
sufficiently small, the inequality $$
\|G(k,z)-G(k,E_{\text{min}}(k))\|< \lambda_{n}-1 $$ holds for any
$\psi\in {\cH}_n$, we have
\begin{equation}\label{relation}
(G(k,z)\psi,\psi)=(G(k,E_{\text{min}}(k))\psi,\psi)-
([G(k,E_{\text{min}}(k))-G(k,z)]\psi,\psi)>(\psi,\psi) .
\end{equation}

From \eqref{relation} we obtain that
\begin{equation}\label{teskari}
 n_{+}(1,G(k,z))\geq \dim \cH_{n} =n_{+}(1,G(k,E_{\text{min}}(k)))
\end{equation}
for any $z\in U_{\delta}(E_{\text{min}}(k)),\delta>0$ sufficiently
small.

 Combining \eqref{monoton} and \eqref{teskari} we obtain
\begin{equation}\label{tenglik}
n_{+}(1,G(k,z))=n_{+}(1,G(k,E_{\text{min}}(k)))\end{equation} for
all $z\in U_{\delta}(E_{\text{min}}(k))$.

By Birman-Schwinger's principle (Lemma \ref{l.b-s})and the
equality \eqref{tenglik} we have
\begin{equation}\label{tenglikH}
\lim_{z\to E_{\text{min}}(k)-0} n_{-}(z,H
(k))=n_{-}(E_{\text{min}}(k),H (k)).
\end{equation}

Taking into account \eqref{tenglik} and \eqref{tenglikH}
 we obtain
the equality \eqref{b-s=}.
\end{proof}
\section{ Threshold analysis of $H(k)$, $k\in
{\bbT}^3$ }

\begin{proposition}\label{ker}
Assume Assumption \ref{hypoth}.
Then the operator $H(0)$ has a nontrivial kernel if and only  if
the integral operator $G(0,0)$ has the eigenvalue $\lambda=1$ and
the corresponding eigenfunction $\psi\in L^2(\T^3)$ satisfies the
condition $$\int_{\T^3}v^\frac{1}{2}(t)\psi(t)dt=0.$$
\end{proposition}
\begin{proof}
See \cite{ALMzMma05}.
\end{proof}

\begin{definition}\label{def}
Let Assumption \ref{hypoth} be fulfilled. The operator $H(0)$ is
said to have a zero energy resonance if the integral operator
$G(0,0)$ has  the eigenvalue $\lambda=1$ and the associated
eigenfunction $\psi$  satisfies the condition
\begin{equation*}
(v^{1/2},\psi)=\int_{\T^3} v^{1/2}(p)\psi(p)dp\not=0.
\end{equation*}
\end{definition}
\begin{remark}
Our definition \ref{def} of a zero energy resonance is a direct
analogue of that
 in the continuous case
 (see, e.g.,  \cite{AGHH}, \cite{Sob}, \cite{Tam93} and
\cite{Yaf79} and references therein).
\end{remark}

\begin{remark}
If the Hamiltonian $H(0)$  has a zero energy resonance
then the
  function
$$
f(p)=\frac{(V^{\frac{1}{2}}\psi)(p)}{E_0(p)} $$  obeys the
equation $$ H(0)f=0$$ and $f$ belongs to $ L^{r}(\bbT^3)$, $1\le
r<3/2$.
\end{remark}

The following lemma expresses an important feat of the
two-particle Schr\"odinger operators.
\begin{lemma}\label{positivity} Let $m_{1}=m_{2}=m$ and let $H(0)$
be positive.  Then for all $k\in \T^3,k\neq 0$ the operator $H(k)$
is positive.
\end{lemma}
\begin{proof} Since $$E_{k}
(p)=\frac{1}{m}\big
(\varepsilon(\frac{k}{2}-p)+\varepsilon(\frac{k}{2}+p) \big )$$
and $v(p)$ are even functions, the
 subspace $L^2_{e}(\T^3)$ resp. $L^2_{o}(\T^3)$ of  the even resp. odd  functions of
${L^2}({\bbT}^3)$ is an
invariant subspace for the operator
 $H(k)$.

Then for any $f \in L^2(\T^3)$ we have
\begin{equation}\label{even+odd}
(H(k)f,f)=(H(k)f_e,f_e)+(H(k)f_o,f_o),
\end{equation}
where $$ f_e(p)=\frac{f(p)+f(-p)}{2},\quad
f_o(p)=\frac{f(p)-f(-p)}{2} .
$$

We remark that $f_e \in L^2_{e}(\T^3) $ and $f_o \in
L^2_{o}(\T^3).$ Therefore making a change of variables on the
r.h.s of \eqref{even+odd} we have
\begin{equation*}
(H(k)f,f)=(H(0)f_{ek},f_{ek})+(H(0)f_{ok},f_{ok})\geq 0\,\,
\mbox{for}\,\, f\neq 0,
\end{equation*}
where $$
 f_{ek}(p)=
f_e(\frac{k}{2}-p),\,f_{ok}(p)= f_o(\frac{k}{2}-p).
$$

We note that $ f_{ek},\,f_{ok}
  \in L^2(\T^3).$
\end{proof}
Our main nonperturbative result is the following
\begin{theorem}\label{existence} Let $m_{1}=m_{2}=m $ and assume
that the operator $H(0)$ is positive. Assume moreover that $H(0)$
has a zero eigenvalue of multiplicity $n$ and a zero energy
resonance. Then for all nonzero $k\in (-\pi,\pi)^3$ the operator
$H(k)$ has at least $n+1$ nonnegative eigenvalues lying below the
bottom $E_{\text{min}}(k)$ of the band spectrum of $H(k).$
\end{theorem}
\begin{proof} Under the assumptions of Theorem \ref{existence} the equation
$G(0,0)\psi=\psi$ has $n+1$ solutions $\psi_0,\psi_1,...,\psi_n$
in the Hilbert space $L^2(\T^3)$. Let $\psi\in L^2(\T^3)$ be one
of the solutions.
 Then
\begin{equation*}
 \int\limits_{{\T}^3}(E_0(p))^{-1}|(V^{1/2}\psi)(p)|^2 d p
 =(G(0,0)\psi,\psi)=(\psi,\psi).
\end{equation*}
For any  nonzero $k\in(-\pi,\pi)^{3}$ and $p\neq 0$ the
inequalities $$0<E_k(p)-E_{\text{min}}(k)<E_0(p)$$ hold.
Therefore, by the definition \eqref{yadroG} of the operator
$G(k,E_{\text{min}}(k))$  we have
\begin{align*}
(\psi,\psi)=\int\limits_{{\T}^3}\frac{|(V^{1/2}\psi) (p)|^2 d
p}{E_0(p)}< \int\limits_{{\T}^3}\frac{|(V^{1/2}\psi) (p)|^2 d
p}{E_k(p)-E_{\text{min}}(k)}= (G(k,E_{\text{min}}(k))\psi,\psi)
\end{align*}
for all nonzero $k\in(-\pi,\pi)^3.$ This means $$
n_{+}(1,G(k,E_{\text{min}}(k)))\geq n+1.
$$ Applying the Birman-Schwinger principle we  conclude that for
all nonzero $k\in(-\pi,\pi)^3$ the operator $H(k)$ has at least
$n+1$ eigenvalues lying below the bottom of the essential
spectrum. Since by Lemma \ref{positivity} the operator $H(k)$ is
positive we have that all eigenvalues of $H(k)$ are nonnegative.

\end{proof}
\begin{corollary}\label{existences} Let $m_{1}=m_{2}=m $ and assume
that the operator $H(0)$ is positive. Assume that $H(0)$ has a
zero eigenvalue of multiplicity $n$. Then for all nonzero $k\in
(-\pi,\pi)^3$ the operator $H(k)$ has at least $n$ nonnegative
eigenvalues lying below the bottom $E_{\text{min}}(k)$ of the band
spectrum of $H(k).$
\end{corollary}

\begin{remark} We note that the interaction operator $V$ is positive
and hence $H(k)$ has no eigenvalue bigger than $E_{max}(k).$
\end{remark}

\begin{remark} A result related to Theorem \ref{existence} has been obtained
in \cite{ALMzMma05}. In the latter reference only existence of
eigenvalues below $E_{\text{min}}(k)$ has been in fact proven,
using a different method, however it was not proven that these
eigenvalues are non negative. See also \cite{ALMzMma05} for
examples, which in particular show the existence of a zero-energy
resonance and zero-eigenvalue of multiplicity 2 or 3.
\end{remark}

\begin{theorem}\label{cheksiz}
Assume  $m_{1}=m_{2}$ and $k^{(j)}=\pi$ for some $j= 1,2,3.$ Then
$w_{jb}(\pi)=0$ and the following inequality
\begin{equation*}
n_{-}(E_{\text{min}}(k),H(k))\geq card\{s\in \Z^1: \hat
v(se^j)>0\}
\end{equation*}
holds.
\end{theorem}
\begin{proof}
Under the assumptions of Theorem \ref{cheksiz} the function
$E_{k}(p)$ does not depend on $p^{(j)}$ and $w_{jb}(\pi)=0.$ Note
that for any $x\in\Z^3$ the value $\hat{v} (x)$ is an eigenvalue
of the operator $V.$ Let $\aleph_{j} =\{s\in \Z^1: \hat
v(se^j)>0\}.$ Then the associated eigenfunctions $$
\psi_{s}(p^{(j)})=(2\pi)^{-3/2}e^{i s p^{(j)}}, s\in \aleph_{j},
$$ of the operator $H(k)$ depend only on $p^{(j)}.$ Let $\cH_n$ be
the $n (n\leq card\,\, \aleph_{j})$-dimensional subspace of the
Hilbert space $L^2(\T^3)$ generated by the eigenfunctions
$\{\psi_{s_{i}}(p^{(j)})\}_{i=1}^{n}, s_{i}\in \aleph_{j}.$ Then
the subspace $\cH_n$ is  invariant with respect to the operator
$V^{1/2}$ and  the equality
 $$
 \inf_{\psi\in \cH_n}\|V^\frac{1}{2}\psi\|^2=
 \inf_{\psi\in \cH_n}
 (V^\frac{1}{2}\psi,V^\frac{1}{2}\psi)=\inf_{\psi\in \cH_n }
 (V \psi,\psi)=\min_{1\leq i\leq n}\hat{v}(s_i e^{j})\|\psi\|^{2}$$
holds.  Now we will prove that for any natural $n \leq card
\aleph_{j}$ there is $z_n< E_{\text{min}}(k)$ such that
\begin{equation}\label{kattabir}
 \inf_{\psi\in \cH_n ,\|\psi\|=1}(G(k,z_{n})\psi,\psi)>1. \end{equation}
 By Birman-Schwinger principle this means
$$ n_{-}(E_{min}(k),H(k))\geq n_{+}(1,G(k,z_n)) \geq  card
\aleph_{j}.$$
 Indeed $$
 \inf_{\psi\in \cH_n,\|\psi\|=1}(G(k,z)\psi,\psi)=
 $$$$\inf_{||\psi||=1, \psi\in \cH_n}\int_{\T^1}|(V^\frac{1}{2}\psi)(p^{(j)})|^2dp^{(j)}
 \int_{\T^2}(E_k(p)-z)^{-1}\frac{dp}{dp^{(j)}}=$$
 $$= \frac{1}{(2\pi)^2}\min_{1\leq i\leq n}\hat{v}(s_i e^{j})
 \int_{\T^2}(E_k(p)-z)^{-1}\frac{dp^{(1)}dp^{(2)}dp^{(3)}}{dp^{(j)}}.$$
 One can check
$$ \lim _{z \to E_{\text{min}}(k)-0}
\int_{\T^2}\frac{dp^{(i)}dp^{(l)}}{E_k(p)-z}= \lim _{z \to
E_{\text{min}}(k)-0}(C_0(k)-\frac{2\pi
\log(E_{\text{min}}(k)-z)}{\sqrt{\cos\frac{k^{(i)}}{2}
\cos\frac{k^{(l)}}{2}}})=+\infty,
 $$ where $i,l=1,2,3,$ $i\neq l, j\neq l, i\neq j $ and $C_0(k)$ is a positive
number. Therefore there exists a number $z_n< E_{\text{min}}(k)$
satisfying inequality \eqref{kattabir}.
\end{proof}
The Theorem \ref{cheksiz} immediately yields the following
\begin{corollary}
Let the  perturbation operator $V$ have infinitely many positive
eigenvalues. Then under assumptions of Theorem \ref{cheksiz} the
operator $H(k)$ has infinitely many eigenvalues below the bottom
$E_{min}(k)$ of the band.
\end{corollary}

{\bf Acknowledgments} The authors very grateful to
Prof.~Dr.~K.~A~.Makarov and Dr. Z.~I.~Muminov for useful
discussions. This work was partially supported by the DFG 436 USB
113/4 Project and the Fundamental Science Foundation of
Uzbekistan. S.~N.~Lakaev and J.~I.~Abdullaev gratefully
acknowledge the hospitality of the Institute of Applied
Mathematics and of the IZKS of the University of Bonn.

\end{document}